# GRN+: A Simplified Generative Reinforcement Network for Tissue Layer Analysis in 3D Ultrasound Images for Chronic Low-back Pain

**Zixue Zeng,[a, b] Xiaoyan Zhao,[a] Matthew Cartier,[c] Xin Meng,[a] Jiantao Pu[a,b,d,*]**

[a]University of Pittsburgh, Department of Bioengineering, Pittsburgh, USA
[b]University of Pittsburgh, Department of Radiology, Pittsburgh, USA
[c]University of Pittsburgh, Department of Mathematics, Pittsburgh, USA
[d]University of Pittsburgh, Department of Ophthalmology, Pittsburgh, USA

**Abstract**

3D ultrasound delivers high-resolution, real-time images of soft tissues, which is essential for pain research. However, manually distinguishing various tissues for quantitative analysis is labor-intensive. To streamline this process, we developed and validated GRN+, a novel multi-model framework that automates layer segmentation with minimal annotated data. GRN+ combines a ResNet-based generator and a U-Net segmentation model. Through a method called Segmentation-guided Enhancement (SGE), the generator produces new images and matching masks under the guidance of the segmentation model, with its weights adjusted according to the segmentation loss gradient. To prevent gradient explosion and secure stable training, a two-stage backpropagation strategy was implemented: the first stage propagates the segmentation loss through both the generator and segmentation model, while the second stage concentrates on optimizing the segmentation model alone, thereby refining mask prediction using the generated images. Tested on 69 fully annotated 3D ultrasound scans from 29 subjects with six manually labeled tissue layers, GRN+ outperformed all other semi-supervised methods in terms of the Dice coefficient using only 5% labeled data, despite not using unlabeled data for unsupervised training. Additionally, when applied to fully annotated datasets, GRN+ with SGE achieved a 2.16% higher Dice coefficient while incurring lower computational costs compared to other models. Overall, GRN+ provides accurate tissue segmentation while reducing both computational expenses and the dependency on extensive annotations, making it an effective tool for 3D ultrasound analysis in cLBP patients..

## 1 Introduction

Chronic lower back pain (cLBP) is a prevalent and complex condition [1], affecting over 70% of individuals in industrialized nations back pain at some point in their lives [2]. Despite extensive research efforts, the understanding of cLBP remained limited, leading to insufficient treatment options [3].

Various imaging modalities, including computed tomography (CT), magnetic resonance imaging (MRI), and ultrasound, have been explored to identify biomarkers for cLBP diagnosis.

While CT and MRI offer high resolution of anatomical structures and their changes, their high cost and accessibility limit their routine use. Ultrasound, by contrast, is affordable, fast, and radiation-free, making it a practical option for cLBP evaluation. Current cLBP research predominantly uses two-dimensional (2D) ultrasound, which has identified distinct characteristics in the thoracolumbar fascia (TLF), multifidus (MF), and transversus abdominis (TrA) [4-8]. However, 2D imaging cannot capture the full 3D characteristics of anatomical tissues. 3D ultrasound overcomes these limitations by enabling precise layer-by-layer analysis of tissue structures from the dermis to muscles. Despite its potential, 3D ultrasound faces challenges, including the need to analyze hundreds of image slices per exam and make manual layer-by-layer annotation both time-consuming and impractical. Therefore, it is greatly desirable to have a tool that can automate the identification of the anatomical tissues depicted on 3D ultrasound images to facilitate cLBP studies.

Few studies have been performed to analyze the anatomical structures depicted on 3D ultrasound images related to cLBP. In machine learning, various methods have been developed to address training artificial intelligence (AI) models with limited annotation data. Semi-supervised learning (SSL) combines supervised training on small labeled datasets with unsupervised learning on larger unlabeled datasets, often using techniques like auxiliary task design, contrastive learning, or interpolation consistency training [9-13]. For instance, the uncertainty-aware mean teacher (UAMT) framework minimizes consistency loss between student and teacher models on unlabeled images [14]. However, SSL often requires extensive unlabeled datasets and relies on the assumption that smoother predictions are more accurate, which is not always valid. An alternative approach involves using complex model architectures with attention mechanisms or deeper convolutional networks to enable sophisticated feature extraction [15-17]. While these models can improve segmentation accuracy, they typically require more diverse datasets and longer training time.

We introduce a simplified and innovative framework called Generative Reinforcement Network+ (GRN+). Unlike the original generative reinforcement network (GRN) [18], which integrates a Generative Adversarial Network (GAN) into the segmentation training process, GRN+ adopts a more simplified strategy by directly concatenating an image generator with a segmentation model (segmentor) into a unified architecture. The generator produces and refines images for the segmentation model, while the segmentation loss is propagated through both components. This approach allows for richer feature extraction and improves segmentation performance without relying on complex structures like attention mechanisms. GRN+ supports sample-efficient learning (SEL) without requiring large unlabeled datasets, as the generator dynamically produces and refines diverse images, thereby effectively expanding the training set for the segmentation model. To ensure stability and prevent gradient explosion, we implement a multi-stage backpropagation strategy. In the first stage, the segmentation loss is propagated through both the generator and the segmentor, while in the second stage, it is confined to the segmentor. This dual-gradient approach stabilizes training and enhances segmentation accuracy. The GRN+ achieves robust performance on both limited annotated datasets and fully annotated datasets, making it a valuable tool for facilitating 3D ultrasound image analysis related to cLBP.

## 2 Materials and Methods

*2.1 Study Dataset*

The dataset was collected as part of an ongoing NIH-funded project (IRB: STUDY22090014) involving participants with cLBP and healthy controls. Participants were positioned face-down on an examination table. The ultrasound array transducer was placed laterally to the midline at the L3-L4 vertebral interspace (see Figure 1, Part 1), focusing on the MF and erector spinae (ES) muscles. A row-column array (RCA) transducer (RC6gV, Vermon) with a central frequency of 6 MHz and an active aperture of 25.6 mm × 25.6 mm was connected to the Vantage 256 ultrasound system (Verasonics Inc., WA, USA) to acquire 3D volumetric ultrasound data. A synthetic aperture technique was utilized to improve image quality. During image acquisition, each transducer element sequentially transmitted a pulse, and all elements simultaneously received echo signals following each transmission. This approach enabled dynamic focusing during beamforming, facilitating high-quality image reconstruction.

Temporal compounding was applied by averaging three consecutive frames to improve the signal-to-noise ratio. Ultrasound scans were performed on both the left and right sides of the back. On each side, an experienced ultrasound examiner targeted two specific locations over the multifidus MF and erector spinae (ES) muscles, with each site scanned three times. This process resulted in a total of 12 B-mode scans per participant.

The study included 29 enrolled participants, from which 69 scans (N_image=17,664) across various locations were randomly selected for algorithm development (see Table 1). An experienced ultrasound operator meticulously annotated six distinct anatomical layers on the 3D ultrasound scans: dermis, superficial fat, superficial fascial membrane (SFM), deep fat, deep fascial membrane (DFM), and MF muscle (refer to Figure 1, Parts 2 & 3). The annotated dataset was divided into training ($N_{image} = 10,722, N_{scan} = 44, N_{patient} = 16$), internal validation ($N_{image} = 512, N_{scan} = 2, N_{patient} = 2$) ), and independent test ($N_{image} = 6,430, N_{scan} = 23, N_{patient} = 11$) sets at the patient level. A larger test set, comprising 23 scans, was allocated to provide a robust assessment of model performance.

**Table 1** Descriptive statistics information for demographic variables and LBP status.

| N = 29 | Summary |
|---|---|
| Demographic variable | |
| Age | 47.11(19.34) |
| Female | 62.07% |
| Hispanic or Latino | 3.44% |
| Black or African American | 6.90% |
| White | 93.10% |
| Height (in inches) | 66.38(3.80) |
| Weight (in pounds) | 179.83(36.02) |
| LBP status | |
| Positive | 72.41% |

Summary statistics are reported as mean (standard deviation) for continuous measurements and percentages for categorical measurements.

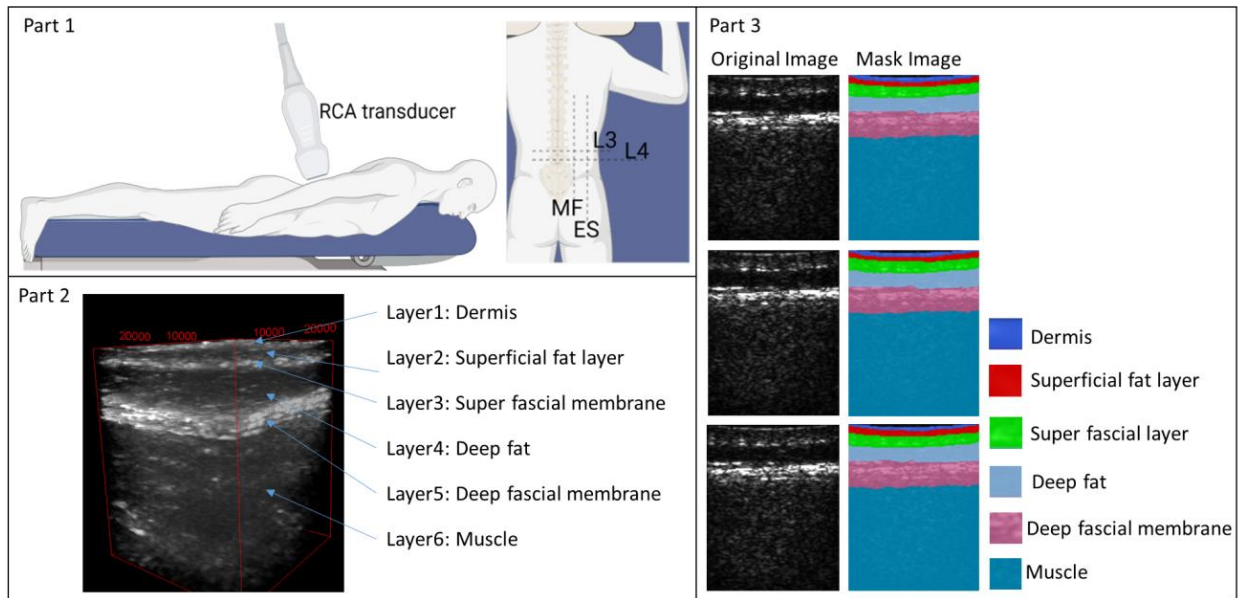

**Fig. 1** 3D ultrasound B-mode image acquisition and annotations. Part 1: Ultrasound image acquisition process. Part 2: A 3-D ultrasound B-mode image showing six distinct tissue layers. Part 3: Original B-mode images along the coronal or sagittal views paired with their corresponding annotations, where each color represents a specific tissue layer.

## 2.2 A Simplified Generative Reinforcement Network (GRN+)

In our original GRN method (upper section, Figure 2), a GAN model is combined with a segmentation model to form a unified training framework. In each training iteration, the generator receives a labeled image and produces a refined version, which is then passed to the segmentor. The segmentor's mask predictions are used to compute the segmentation loss, which is backpropagated to the generator for improvement. Additionally, the generator is trained using a combination of segmentation loss ($L_{seg}$), pixel difference loss ($L_{pixel}$), and discriminator loss ($L_{Disc}$):

$$min_G(\alpha L_{seg} + \beta L_{pixel} + \gamma L_{Disc}) \tag{1}$$

where $G$ represents the generator, and $\alpha$, $\beta$ and $\gamma$ are hyperparameters designed to balance each loss term. While the generator aims to create realistic images and minimize segmentation loss, this dual objective can compromise segmentation accuracy, as improving image realism may conflict with enhancing segmentation performance. To address this limitation, we developed a novel architecture, GRN+, where the generator is optimized solely based on segmentation loss feedback.

$$min_G \left(L_{seg}(S(G(x)), y)\right), where\ x \sim p_{data}(x) \tag{2}$$

where x is sampled from the data distribution $p_{data}(x)$, y is the ground truth, $G$ is the generator, and $S$ is the segmentor. The generator in GRN+ is specifically designed to produce enhanced images that directly reduce segmentation loss, without relying on L1 loss or discriminator loss. This allows the generator to adjust pixel values to emphasize regions of interest or eliminate irrelevant details in the images. During each training iteration, the segmentation loss is backpropagated directly to the generator, enabling iterative refinement of the generated images and improving segmentation performance without competing objectives.

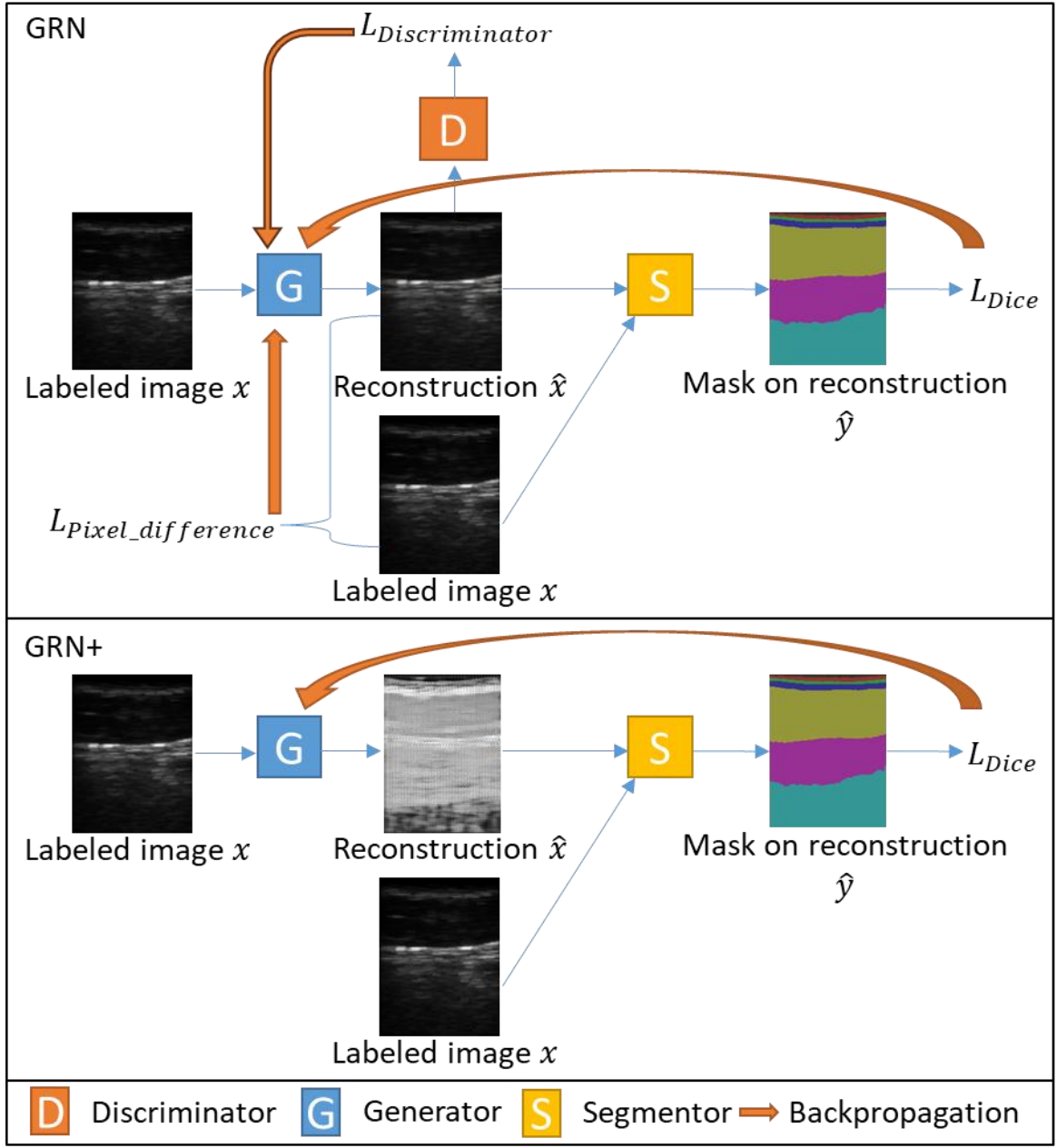

**Fig. 2** Comparison of GRN and GRN+. In GRN, the generator is trained using three loss functions: $L_{Discriminator}, L_{Dice}, L_{pixel\_difference}$. In contrast, GRN+ simplified the segmentation feedback mechanism by optimizing the generator solely based on the segmentation loss $L_{Dice}$, allowing for more direct and efficient training.

*2.2.1 GRN+ with Multi-stage Backpropagation*

The proposed GRN+ framework integrates the generator and segmentor into a unified model. Let $G(\cdot, \theta_G)$ represents the generator with the parameter $\theta_G$ and $S(\cdot, \theta_S)$ the segmentor with parameter $\theta_S$. The original image $x$ is first passed through the generator to produce a new image $\hat{x}$, which is then input into the segmentor to generate the mask prediction $\hat{y}$. The mask prediction is expressed as:

$$\hat{y} = S(G(x, \theta_G), \theta_S) \tag{5}$$

To stabilize the training of the concatenated multi-model structure, we introduce a multi-stage backpropagation mechanism (Figure 3). In Stage 1, the original image is input into the concatenated model $S(G(x, \theta_G), \theta_S)$ to produce a mask prediction $\hat{y}$. A segmentation loss is calculated and backpropagated through both the generator and the segmentor.

$$L_{seg}^{(1)} = L_{seg}(S(G(x, \theta_G), \theta_S), y) \tag{6}$$

In Stage 2, the segmentation model takes both the original image $x$ and the newly reconstructed images $\hat{x}$ as inputs to calculate a combined segmentation loss, which is then backpropagated to the segmentor to enhance its robustness, ensuring that it can make independent predictions without relying on the generator.

$$L_{seg}^{(2)} = L_{seg}(S(G(x, \theta_G), \theta_S), y) + L_{seg}(S(x, \theta_S), y) \tag{7}$$

The motivation for multi-stage backpropagation is driven by the complexity of concatenating multiple models, which increases the overall depth of the unified architecture. This added depth can lead to gradient explosion during backpropagation as gradients propagate through several intricate encoder-decoder structures [19]. Based on the chain rule, the gradient of $\theta_G$ with respect to $L_{seg}$ is:

$$\nabla_{\theta_G} L_{seg} = \underbrace{\frac{\partial L_{seg}}{\partial \hat{y}}}_{\text{can be large if } \hat{y} \text{ is poor}} \times \underbrace{\frac{\partial \hat{y}}{\partial \hat{x}}}_{\text{depend on } \theta_S} \times \underbrace{\frac{\partial \hat{x}}{\partial \theta_G}}_{\text{depend on } \theta_G} \tag{8}$$

This loss backpropagation can potentially lead to a gradient explosion problem due to the following cycle:

1. Incorrect mask predictions lead to unstable gradient updates for $\theta_G$.

2. Since the segmentor uses the newly reconstructed image $\hat{x}$, unstable $\theta_G$ could result in poor inputs for the segmentor and inaccurate mask prediction $\hat{y}$.

3. The degraded mask prediction $\hat{y}$ increases $L_{seg}$, which in turn exacerbates instability.

By incorporating Stage 2, where only the segmentor is optimized while the generator's weights are frozen, the segmentation loss is minimized without inducing large magnitude gradient updates in the generator. This reduces the magnitude of the segmentation loss and thereby stabilizes the segmentation gradient with respect to the mask prediction $\frac{\partial L_{seg}}{\partial \hat{y}}$. Additionally, updating the segmentor's weights twice per iteration accelerates the convergence of the segmentor model weights $\theta_S$, which further reduce the magnitude of $\frac{\partial \hat{y}}{\partial \hat{x}}$. Consequently, the norm of the gradient of $\theta_G$ with respect to the segmentor loss becomes:

$$\|\nabla_{\theta_G} L_{seg}\| = \left\| \underbrace{\frac{\partial L_{seg}}{\partial \hat{y}}}_{lower\ because\ of\ stage\ 2\ backpropagation} \times \underbrace{\frac{\partial \hat{y}}{\partial \hat{x}}}_{stabilize\ by\ faster\ S\ convergence} \times \underbrace{\frac{\partial \hat{x}}{\partial \theta_G}}_{depend\ on\ \theta_G} \right\| \quad (7)$$

Since both $\frac{\partial L_{seg}}{\partial \hat{y}}$ and $\frac{\partial \hat{y}}{\partial \hat{x}}$ are reduced and stabilized in Stage 2, the magnitude of $\nabla_{\theta_G} L_{seg}$ remains bounded. This constrained gradient update reduces the segmentation loss during training, stabilizing the generator's weight updates and effectively mitigating the gradient explosion problem.

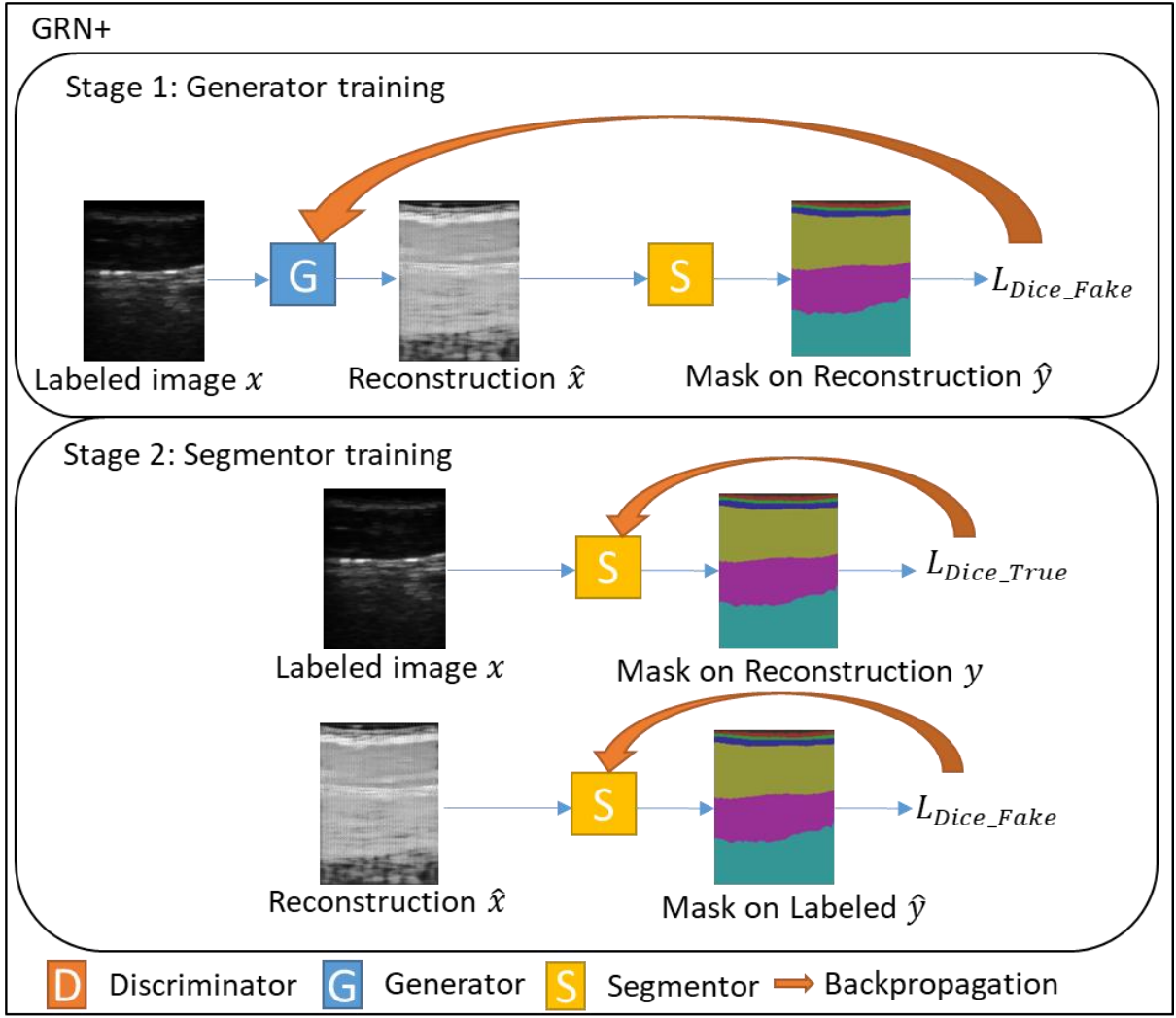

**Fig. 3** Workflow of GRN+ with multi-stage backpropagation. In Stage 1, the segmentation loss is backpropagated to the generator, updating the weights of both the segmentor and the generator. In Stage 2, the segmentor receives both the newly reconstructed and original images as inputs and updates its weight based on predictions from these two image types. This multi-stage backpropagation mechanism accelerates the weight updates for the segmentor while stabilizing the generator, effectively preventing the gradient explosion issue.

---

**Algorithm Overview** Generative Reinforcement Network + (GRN+)

**Require:** $G_\theta$: Generator with trainable parameters $\theta$
**Require:** $S_{\theta'}$: Segmentor with trainable parameters $\theta'$
**Require:** $Dataset_L(I, M)$: Collection of labeled samples
**Require:** $T$: total number of iterations
    **for** $t = 1, ..., T$ **do**
        Sample $I_L, M \sim Dataset_L(I, M)$
        $\widehat{I_L} = G_\theta(I_L)$
        $\widehat{M_L} = S_{\theta'}(\widehat{I_L})$
        $L_G = \lambda_{seg} L_{Dice}(\widehat{M_L}, M_L)$

    **Update** $\theta, \theta'$ based on $L_G$
    $M_L = S_{\theta'}(I_L)$
    $L_S = (L_{Dice}(M_L, M), L_{Dice}(\widehat{M_L}, M))/2$
    **Update** $\theta'$ based on $L_S$
  **end for**
  **return** $\theta, \theta'$

---

*2.2.2 Dynamic Generator Augmentation (DGA) and Segmentation-guided Enhancement (SGE)*

Similar to GRN, GRN+ incorporates dynamic generator augmentation (DGA). During each training iteration, the generator's weights are updated iteratively using the formula:

$$\theta_{t+1} \leftarrow \theta_t - \eta \cdot \nabla_{\theta_t} L_{seg}^{(1)} \tag{8}$$

where $\eta$ is the learning rate, and $\nabla_{\theta_t}$ is the gradient calculated at iteration $t$. Even with identical input images, the generator's reconstructed output differs due to continuous weight updates $G_{\theta_{t+1}}(I) \neq G_{\theta_t}(I)$. This dynamic process improves the segmentor's robustness by training it on a diverse range of dynamically generated images, making it especially useful for scenarios with limited dataset sizes.

In the GRN+ framework, since the generator is optimized to produce enhanced reconstructed images that minimize the segmentor loss $min_G \left( L_{seg}(S(G(x)), y) \right)$, the generator and segmentor can be combined as a unified model during inference. This yields a joint prediction $y = S(G(x))$, where the segmentor receives pre-enhanced images from the generator for mask prediction. We refer to this sequential approach as Segmentation Guided Enhancement (SGE) and introduce two variants of the GRN+ method. The first variant, the vanilla GRN+ method, uses the segmentor alone for mask predictions. The second variant, GRN+ with SGE, utilizes the concatenated generator-segmentor architecture for mask prediction.

*2.3 Backbone Model Architecture*

The core segmentation network is a 2D U-Net model with an encoder channel of (16, 32, 64, 128, 256). It is designed for single-channel input and generates segmentation masks for seven classes (six tissue layers and 1 background). The generator's encoder (upper section, Figure 4) incorporates multiple residual blocks to improve gradient flow efficiency and support deeper network layers. The encoding process starts with a convolution layer that reduces spatial dimensions while increasing feature depth, followed by a series of residual blocks that further downsample the input data. In contrast, the decoder uses 2D transposed convolutions to upscale high-dimensional feature representations, effectively reconstructing the images.

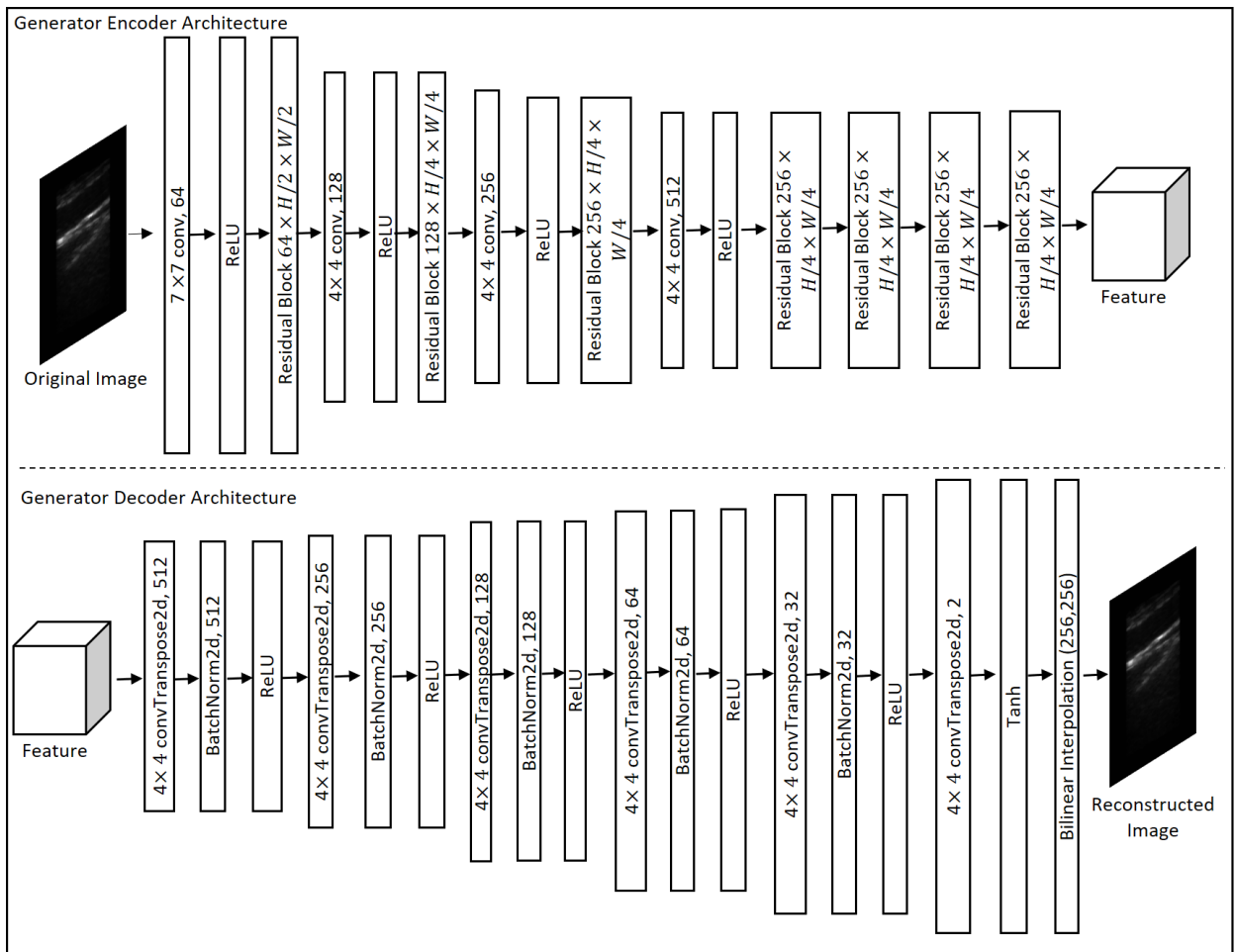

**Figure 4**. Encoder and Decoder Architecture of Generator. The encoder (upper section) utilizes multiple residual blocks to progressively capture high-dimensional patterns. The decoder (lower

section) reconstructs the images from these high-level representations using transposed convolution layers, facilitating the accurate generation of segmentation masks.

*2.4 Model training and performance evaluation*

To validate GRN+ on datasets with limited annotations, we designate 5%, 10%, and 20% of the training dataset as labeled subsets, treating the remainder as unlabeled. For comparison, we evaluated seven other SSL methods, including mean teacher (MT)[20], UAMT [14], Uncertainty Rectified Pyramid Consistency (URPC) [21], Shape-aware Semi-supervised network (SASSNet) [22], Regularized dropout (R-drop) [23], Interpolation Consistency Training (ICT) [12] and Deep Adversarial Network (DAN) [24]. Additionally, we compare GRN+ with SGE against the latest semantic segmentation approaches featuring deeper encoder-decoder architectures. These include DeepLabv3+ [25], Swin-Unet [26], SegResNet [27], V-Net [28], as well as the baseline Unet model [29]. The goal was to evaluate whether the GRN+ architecture improve feature extraction compared to existing methods.

All models were implemented using PyTorch and trained on an NVIDIA GTX3090 GPU. The Adam optimizer was utilized to train the generator, segmentor, and discriminator, with an initial learning rate of 0.0002 and exponential decay rates ($\beta_1$, $\beta_2$) set to (0.5, 0.999). The batch size was set to be 4. The training ran up to 50 epochs. The model with the lowest Dice loss on the validation dataset was selected for final evaluation. An early stopping mechanism was implemented, halting training if validation performance did not improve for five consecutive epochs. Segmentation performance was measured using the Dice Similarity Coefficient (DSC) on independent test sets. A higher DSC indicates better alignment between the predicted and the ground truth masks.

## 3 Results

*3.1 Performance evaluations against other Semi-supervised learning methods*

Table 2 presents the performance comparison between the proposed GRN+ method and existing SSL approaches on the independent test dataset under different scenarios. When trained on a dataset with only 5% labeled images, GRN+ significantly outperformed all other SSL methods (p<0.05), even without utilizing unlabeled data for unsupervised training. When training on 10% and 20 % of the labeled images, GRN+ achieved comparable performance to other SSL methods, with differences not statistically significant. However, GRN+ with SGE outperformed all other SSL methods at both 10% and 20% labeled levels.

**Table 2**. Comparison of segmentation performance between our methods and existing SSL methods on the same independent test set.

| Method | Labeled % | | |
| --- | --- | --- | --- |
| | 5% | 10% | 20% |
| Fully supervised | 43.97 (43.71, 44.23) | 64.87 (64.57, 65.17) | 74.52 (74.23, 74.81) |
| MT[20] | 65.09 (64.78, 65.40) | 72.78 (72.47, 73.09) | 76.82 (76.57, 77.07) |
| UAMT[14] | 64.39 (64.04, 64.74) | 72.75 (72.42, 73.08) | 76.95 (76.69, 77.21) |
| URPC[21] | 56.46 (56.17, 56.75) | 73.62 (73.43, 73.81) | 76.94 (76.76, 77.12) |
| SASSNet[22] | 52.66 (52.38, 52.94) | 67.27 (67.02, 67.52) | 76.49 (76.29, 76.69) |
| R-drop[23] | 65.92 (65.57, 66.27) | 75.34 (75.08, 75.60) | 77.35 (76.99, 77.71) |
| ICT[12] | 66.07 (65.75, 66.39) | 74.06 (73.81, 74.31) | 76.54 (76.32, 76.76) |
| DAN[24] | 64.56 (64.30, 64.82) | 73.00 (72.73, 73.27) | 74.34 (74.08, 74.60) |
| GRN+ | **66.36 (66.04, 66.68)** | 71.35 (70.98, 71.72) | 75.26 (74.96, 75.56) |
| GRN+ (w SGE) | 62.31 (61.89, 62.73) | **75.69 (75.36, 76.02)** | **77.92 (77.57, 78.27)** |

Model performance is presented as Dice coefficient (95% confidence interval). Bold denotes the proposed methods that significantly outperform all other SSL methods (p-value < 0.05, paired t-test).

*3.2 Performance evaluation against other segmentation models*

Table 3 presents the comparison of the proposed GRN+ methods and other segmentation models when trained on a fully labeled dataset (100%) and evaluated on the independent test set. The GRN+ model significantly outperformed other segmentation models on 2 out of 6 layers, while GRN+ with SGE achieved superior performance in 4 out of 6 tissue layers (p < 0.05). GRN+ w

SGE outperforms all other methods, achieving a 2.16% high Dice coefficient compared to the best-performing alternative.

**Table 3.** Performance comparison between the proposed GRN+ methods and other segmentation models on the independent test set when trained on all labeled cases in the training set.

| Methods | Dermis | Superficial Fat Layer | SFM | Deep Fat | DFM | Muscle | Overall |
|---|---|---|---|---|---|---|---|
| Unet | 55.73 (55.35, 56.11) | 53.99 (53.65, 54.33) | 73.61 (73.38, 73.84) | 80.90 (80.39, 81.41) | 67.58 (67.26, 67.90) | 87.67 (87.45, 87.89) | 80.62 (80.47, 80.77) |
| Deeplabv3plus | 77.14 (76.94, 77.34) | 64.85 (64.56, 65.14) | 84.29 (84.14, 84.44) | 87.01 (86.48, 87.54) | 81.91 (81.65, 82.17) | 93.42 (93.27, 93.57) | 81.44 (81.29, 81.59) |
| Swin-Unet | 68.50 (68.03, 68.97) | 68.79 (68.42, 69.16) | 85.36 (85.20, 85.52) | 87.93 (87.43, 88.43) | 80.93 (80.66, 81.20) | 93.27 (93.13, 93.41) | 80.80 (80.60, 81.00) |
| V-net | 74.44 (74.12, 74.76) | 72.64 (72.34, 72.94) | 85.61 (85.44, 85.78) | 85.70 (85.21, 86.19) | 81.52 (81.23, 81.81) | 91.62 (91.45, 91.79) | 81.92 (81.76, 82.08) |
| SegResNet | 77.95 (77.68, 78.22) | 72.79 (72.49, 73.09) | 86.93 (86.78, 87.08) | 87.76 (87.27, 88.25) | 76.61 (76.31, 76.91) | 90.96 (90.82, 91.10) | 82.17 (82.03, 82.31) |
| GRN+ | 72.22 (71.93, 72.51) | 71.95 (71.66, 72.24) | 84.65 (84.50, 84.80) | 86.14 (85.65, 86.63) | **82.26 (82.04, 82.48)** | 93.62 (93.46, 93.78) | 81.81 (81.68, 81.94) |
| GRN+ w SGE | 76.48 (76.22, 76.74) | **74.67 (74.37, 74.97)** | 86.33 (86.17, 86.49) | **88.27 (87.79, 88.75)** | 84.98 (84.77, 85.19) | **95.24 (95.09, 95.39)** | **84.33 (84.20, 84.46)** |

Segmentation model performance is presented as Dice coefficient (95% confidence interval). Bold denotes the segmentation performance that statistically outperforms all other segmentation model performance. (p-value < 0.05, paired t-test)

Table 4 compares the proposed GRN+ with the SGE model against other segmentation models with advanced encoder-decoder architectures, evaluating model complexity, computational cost, and segmentation performance. GRN+ with SGE achieved significantly superior segmentation performance compared to other models, as measured by the Dice coefficient (p-value < 0.05). Additionally, GRN+ with the SGE model has fewer model parameters and faster inference time per batch compared to Deeplabv3plus, Swin-Unet, and V-Net models. Notably, the GRN+ variant outperformed the original GRN model in segmentation accuracy while reducing training time per batch by 30%. Figure 5 illustrates the segmentation performance of the proposed GRN+ methods compared with other segmentation models in terms of model complexity.

**Table 4.** Comparison of model complexity, computational cost, and segmentation performance between our method and existing methods when trained on all labeled cases in the training set.

| Method | Parameter Count | Training Time per Batch | Inference Time per Batch | Dice Coefficient |
|---|---|---|---|---|
| Unet | 661727 | 0.0070 | 0.0015 | 80.62 (80.47, 80.77) |
| Deeplabv3plus | 39635271 | 0.0578 | 0.0295 | 81.44 (81.30, 81.58) |
| Swin-Unet | 6302353 | 0.0647 | 0.0198 | 80.80 (80.60, 81.00) |
| V-net | 9362790 | 0.0159 | 0.0127 | 81.92 (81.77, 82.07) |
| SegResNet | 1576487 | 0.0179 | 0.0025 | 82.17 (82.03, 82.31) |
| GRN | 661727(6494152) | 0.0458 | 0.0015 | 80.53 (80.39, 80.67) |
| GRN w SGE | 4450752(2705127) | 0.0458 | 0.0054 | 82.45 (82.33, 82.57) |
| GRN+ | 661727(3789025) | 0.0329 | 0.0015 | 81.81 (81.68, 81.94) |
| GRN+ w SGE | 4450752 | 0.0329 | 0.0054 | **84.33 (84.20, 84.46)** |

Segmentation model performance is presented as Dice coefficient (95% confidence interval). Bold values denote the segmentation performance that is statistically superior to all other model performances (p-value < 0.05, paired t-test). Brackets in the Parameter Count column indicate the additional number of model parameters required exclusively during the training phase. The batch size is set to 4 for all models.

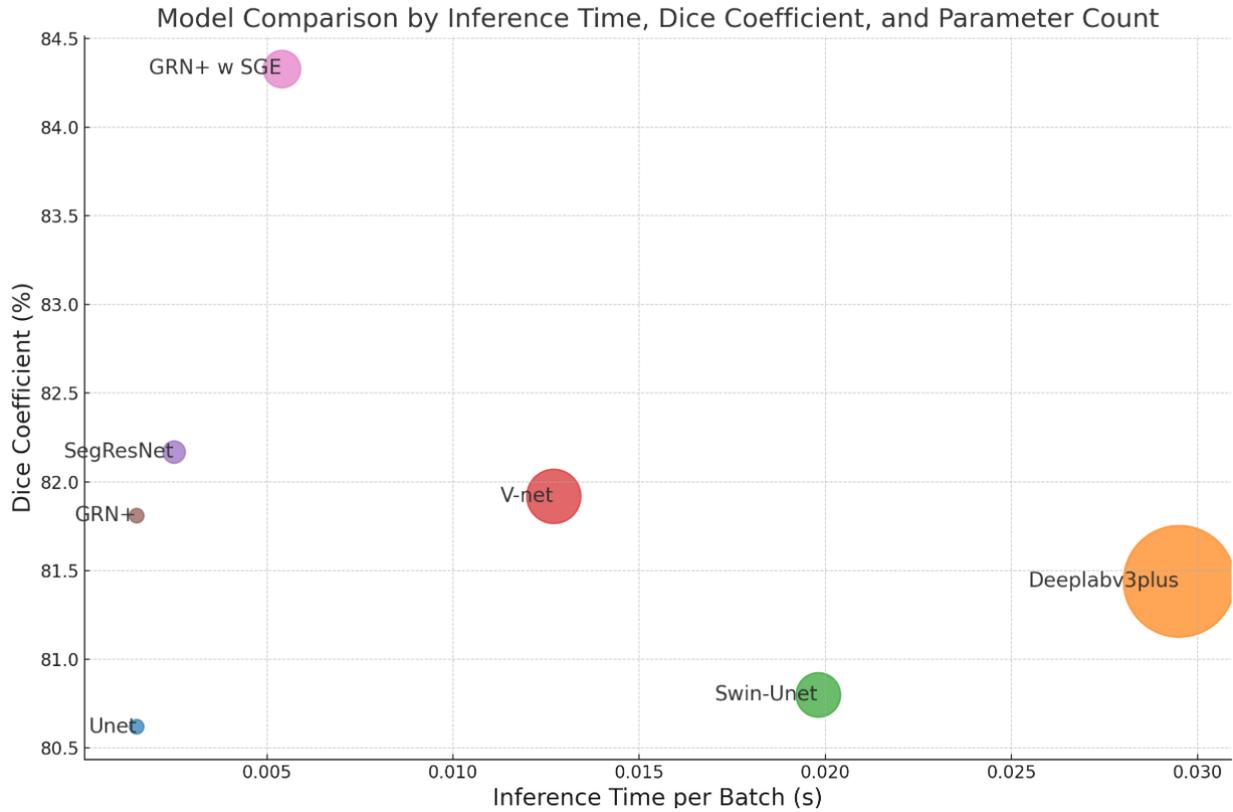

**Figure 5**. Comparison of segmentation performance relative to model complexity. GRN+ with SGE achieves the highest Dice coefficient with only a slight increase in parameter count. All models were evaluated with a batch size of 4.

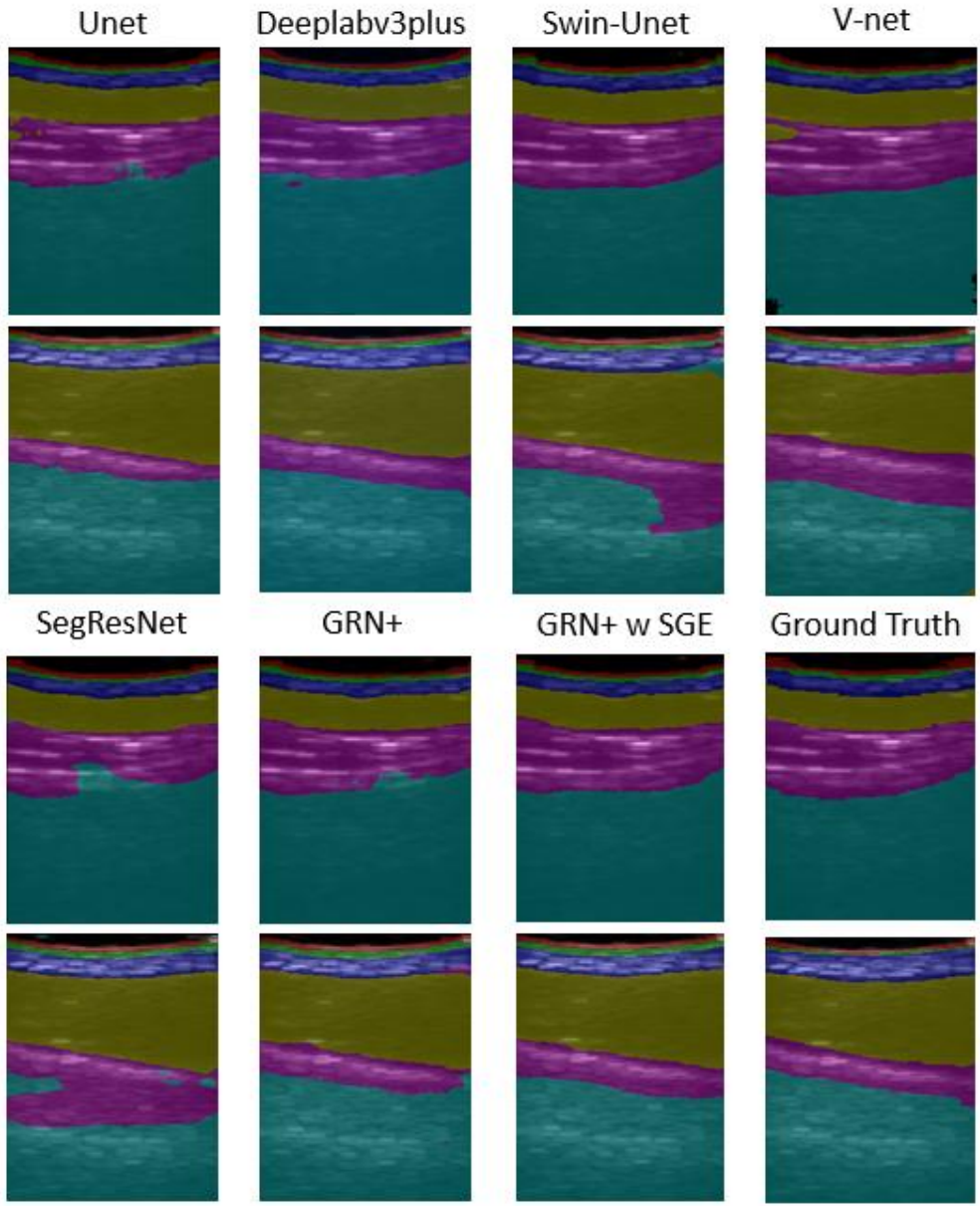

**Figure 6.** Mask prediction for GRN+ model and other segmentation models when trained on 100% labeled dataset. The proposed GRN+ methods demonstrate superior accuracy in tissue layer segmentation compared with other methods.

*3.3 Analysis of SGE and Multi-stage backpropagation*

Figure 6 illustrates the impact of incorporating SGE within the GRN+ framework. Using the reconstructed image as input for the segmentation model significantly improves tissue layer segmentation accuracy. The reconstructed images refine the pixel distribution by brightening dark regions and emphasizing layer boundaries, closely aligning with the ground truth mask. Figure 7 compares the training and validation loss trends for the GRN+ method using single-stage and multi-stage backpropagation. Multi-stage backpropagation substantially reduces both training and validation losses compared to single-stage backpropagation, leading to more stable and efficient convergence. In contrast, single-stage backpropagation demonstrates no clear signs of convergence, highlighting its limitations.

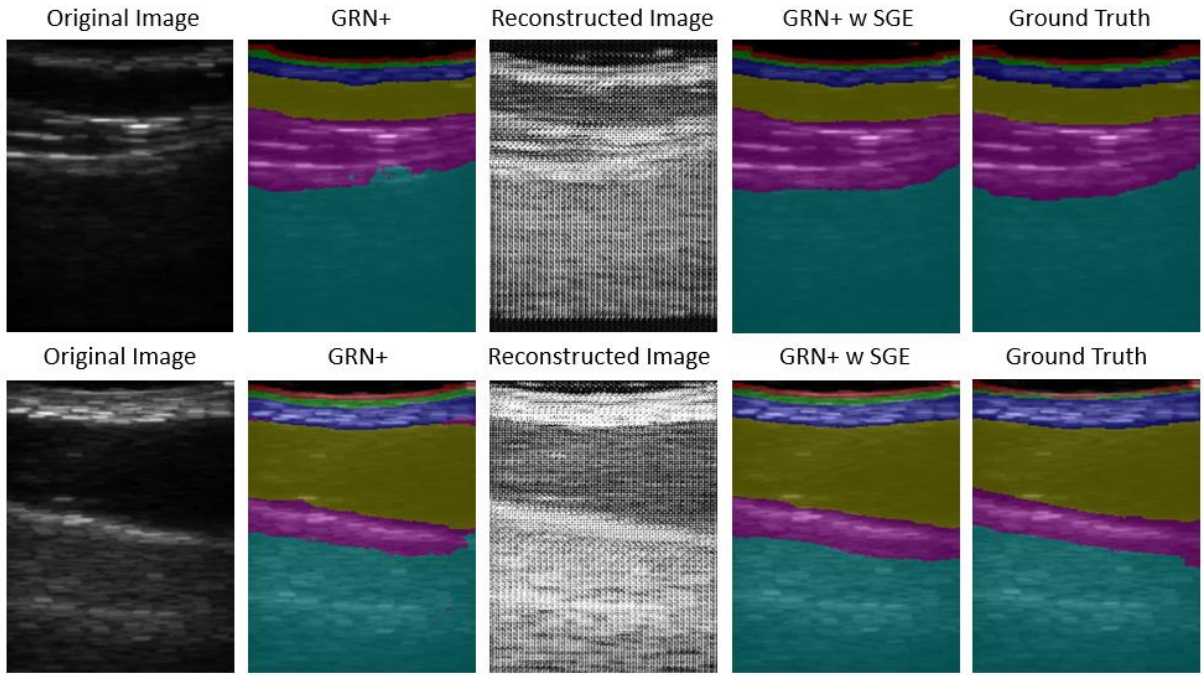

**Figure 7.** Comparison of mask predictions for original and reconstructed images. Using generator-enhanced images results in more accurate tissue layer segmentation.

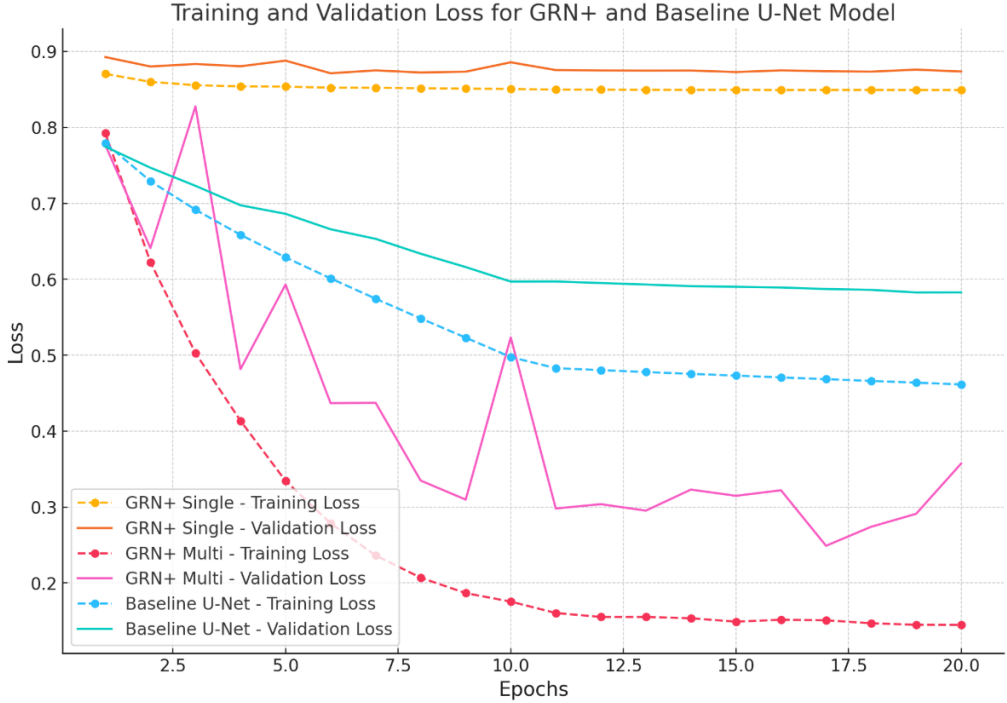

**Figure 8.** Training and validation losses plotted again the number of epochs. "GRN+ Single" refers to the GRN+ model using single-stage backpropagation (without stage 2), while "GRN+ Multi" refers to the model utilizing multi-stage backpropagation.

## 4 Discussion

We propose a novel method called GRN+ that combines a generator and a segmentor in a concatenated multi-model structure to effectively segment tissue layers in 3D ultrasound B-mode images. Unlike traditional encoder-decoder models that increase the model depth and complexity or use attention mechanisms to improve feature extraction, GRN+ integrates a ResNet-based generator before the segmentor. The generator enhances segmentation by preprocessing images to highlight regions of interest and directly adjusting pixel value distribution. Our results show that GRN+ outperforms other SSL approaches when trained on only 5% of labeled datasets without relying on unlabeled images for unsupervised training (Table 2). GRN+ also achieves the highest Dice coefficient with a minimal increase in model complexity, outperforming other segmentation models (Table 4). These findings suggest that GRN+'s effectiveness in segmenting tissue layers with significantly fewer annotated datasets, addressing the challenge of sample-efficient learning.

Unlike the original GRN model, GRN+ eliminates the need for the generator to produce images resembling the original input. Instead, the generator adjusts pixel distributions, enhancing regions of interest with low intensity to improve segmentation. To address gradient explosion, we introduce multi-stage backpropagation, where the segmentor's weights are updated twice during training, allowing faster convergence and reducing gradient oscillations by stabilizing the segmentor's weight updates.

The impact of SGE on segmentation performance depends on the training dataset size. As shown in Tables 2 and 4, SGE significantly improves layer segmentation performance. This improvement is attributed to the increased model complexity through the addition of a ResNet-based generator, allowing the model to learn more complex features. Additionally, the generator produces reconstructed images with adjusted pixel distributions (Figure 7), rescaling dark regions to more moderate pixel values, which accelerates model convergence, as the magnitude of gradient updates is larger for moderate pixel values compared to those near zero. However, with only 5% labeled level, SGE leads to a 4.05% decrease in Dice coefficient, while the GRN+ model without SGE significantly outperforms all other SSLs. This can be attributed to the fact that more complex models require larger datasets for effective training. A dataset with only 5% labeled data does not provide sufficient information for the concatenated model to learn the underlying patterns effectively. In contrast, the generator dynamically produces reconstructed images in each training iteration, significantly increasing dataset diversity. This enhanced diversity strengthens the training of the segmentor, allowing it to outperform all other SSL methods even without unsupervised training.

Compared to more complex segmentation models like Deeplabv3plus, V-Net, and Swin-Unet, GRN+ achieves significantly higher tissue layer segmentation accuracy while requiring fewer parameters and less training time (Table 4). In a single encoder-decoder-based model, the architecture is tasked with multiple responsibilities: detecting features in dark or poorly visible regions, highlighting regions of interest, filtering out irrelevant information or noise, and extracting features for segmentation. This combined workload can overwhelm a single model, particularly when the dataset is small. In contrast, a multi-model setup distributes these tasks between the segmentor and generator, easing the burden on each component for pattern learning. Additionally, both the generator and segmentor contribute to feature extraction, with the generator handling preprocessing and the segmentor performing segmentation. If one model fails to detect certain features, the other can compensate, thereby reducing the overall error rate. This demonstrates that a multi-model approach enhances layer segmentation accuracy more effectively than merely escalating the complexity of a single network.

Our study has limitations. First, due to ongoing patient enrollment and data collection, the dataset remains relatively small. To ensure a robust evaluation of model performance, we retained a large independent test set, comprising 6,430 images derived from 23 scans taken at different locations across 11 subjects. The small dataset size actively motivates the development of our algorithm that can work on the small labeled dataset. Second, with SGE, GRN+ does not outperform other SSL methods at the 10% and 20% labeling levels. However, since GRN+ does not rely on unlabeled data for unsupervised training, its performance has the potential for further improvement. Lastly, the multi-stage backpropagation process increases the training time per batch due to an additional backpropagation step. However, this does not add computational cost during inference, where GRN+ maintains strong performance and minimal inference time.

## 5 Conclusion

We have developed GRN+, a concatenated generator-segmentor multi-model framework tailored for segmenting 3D ultrasound tissue layers in the diagnosis of chronic low back pain (cLBP). To address the issue of gradient explosion, we introduced a multi-stage backpropagation strategy, which separately updates the segmentor model weights to improve segmentation accuracy using generator output. This approach accelerates convergence while enabling joint model training. Our model shows that GRN+ outperforms more complex encoder-decoder architectures when trained on fully labeled data, while also maintaining relatively low computational costs. Unlike semi-supervised learning methods, GRN+ does not depend on unlabeled data for unsupervised training and can effectively utilize minimal labeled data to achieve robust segmentation performance.

*Acknowledgments*

This work is supported in part by research grants from the National Institutes of Health (NIH) (R61AT012282, R01HL174570 and R01CA237277).